\documentclass[aps,prd,preprint,groupedaddress,floatfix]{revtex4}
\usepackage{graphicx}

\newcommand{\nn}{\nonumber}

\newcommand{\pslash}{p\kern-1ex /}
\newcommand{\lslash}{l\kern-1ex /}
\newcommand{\sslash}{s\kern-1ex /}
\newcommand{\Dslash}{{\cal D}\kern-1.5ex /}

\newcommand{\tr}{{\rm tr}}

\newcommand{\beqa}{\begin{eqnarray}}
\newcommand{\eeqa}{\end{eqnarray}}

\begin{document}


\title{
Chiral perturbation theory with Wilson-type fermions including $a^2$ effects:
$N_f=2$ degenerate case
}
\author{$^{1}$Sinya Aoki}
\affiliation{
$^1$Institute of Physics, University of Tsukuba, Tsukuba 305-8571, Japan 
}

\date{\today}

\begin{abstract}
We have derived the quark mass dependence of $m_{\pi}^2$, $m_{\rm AWI}$ 
and $f_{\pi}$, using the chiral perturbation theory which includes the 
$a^2$ effect associated with  the explicit chiral symmetry breaking of the 
Wilson-type fermions, in the case of the $N_f=2$ degenerate quarks.
Distinct features of the results are
(1) the additive renormalization for the mass parameter $m_q$ 
in the Lagrangian,
(2) $O(a)$ corrections to the chiral log ($m_q\log m_q$) term,
(3) the existence of more singular term, $\log m_q$, generated by
$a^2$ contributions,
and
(4) the existence of both $m_q\log m_q$ and $\log m_q$ terms in the
quark mass  from the axial Ward-Takahashi identity, $m_{\rm AWI}$.
By fitting the mass dependence of $m_\pi^2$ and $m_{\rm AWI}$,
obtained by the CP-PACS collaboration for $N_f=2$ full QCD simulations,
we have found that the data are consistently described
by the derived formulae.
Resumming the most singular terms $\log m_q$, 
we have also derived the modified formulae, which show a better control over
the next-to-leading order correction.
\end{abstract}

\maketitle

%
\section{Introduction}
One of the most serious systematic uncertainties in the current lattice
QCD simulations is caused by the chiral extrapolation. Due to the limitation
of the current computational power, one can not perform simulations
directly at the physical light quark(up and down) mass. Instead, one has
performed simulations at several heavier quark masses and has extrapolated 
results to the physical quark mass point, using the polynomial(linear, 
quadratic, etc.) or the formula derived from the chiral perturbation 
theory (ChPT)\cite{ChPT}. 
These extrapolations cause large systematic uncertainties,
in particular in the case of full QCD simulations, where the lightest
quark mass employed in the current QCD simulations is roughly half of 
the physical strange quark mass ($m_\pi/m_\rho \simeq 0.6$).

Recently more serious problem has been pointed out, in particular for 
full QCD simulations with Wilson-type quarks: 
the expected chiral behaviour predicted by the ChPT has not been observed.
For example,  the behaviour of the pion mass $m_\pi^2$
as a function of quark mass $m_q$ is given by
\beqa
m_\pi^2 &=& A m_q [1+\frac{ A m_q}{16\pi^2 N_f f_\pi^2}\log (A m_q/\Lambda^2)],
\eeqa
where $\Lambda$ is some scale parameter. Since the pion decay constant
is experimentally known as $f_\pi = 93$ MeV, only $A$ and $\Lambda$ are unknown
parameters. Unfortunately, such a two parameter fit can not explain lattice 
data well, which looks almost linear in the simulated range of quark masses.
If one includes $f_\pi$ as a free parameter, the best fit typically gives 
$f_\pi^2 \ge 5 \times$ (93 MeV)$^2$\cite{panel}.

The most widely accepted interpretation for this discrepancy
is that the simulated range of quark masses in the current simulations is 
still too heavy to apply the ChPT.
If this interpretation is true, the current lattice simulations with
the (Wilson-type) dynamical quarks lose a large part of
their powers to predict properties of hadrons at the physical
light quark masses.

In this paper, we investigate a theoretically more natural alternative that
the explicit breaking of the chiral symmetry by the Wilson-type quark
action modifies the formulae of the ChPT at the finite lattice spacing.
We first derive formulae in the modified chiral perturbation theory for
the Wilson-type quark action, denoted by WChPT in this paper.
Such attempts have been made before at the leading order\cite{SS} and
the next-to-leading order\cite{RS}. At the leading order\cite{SS}, the WChPT
predicts the existence of the parity-flavor breaking phase 
transition\cite{aoki1,aoki2,aoki3} for the 2 flavor QCD as long as 
massless pions
appear at the critical quark mass. This analysis has also shown that
the $O(a^2)$ chiral breaking term play an essential role to generate
the parity-flavor breaking phase transition, which is necessary to explain
the existence of the massless pions for the Wilson-type quark 
action\cite{aoki1,aoki2,aoki3}. In the next-to-leading order analysis\cite{RS}, however,
only the $O(a)$ breaking effects are included, and it is concluded that
the effect of the chiral symmetry breaking can always be absorbed in the
redefinition of the quark mass, so that all formulae in the ChPT remain the
same if one replaces the quark mass $m_q$ with $m_q-m_c$, where $m_c$ is the
additive $O(a)$ counter-term for the quark mass.
In the section \ref{sec:wcpt}, we perform the next-to-leading order calculation
in the WChPT including $O(a^2)$ chiral symmetry breaking effects.
To make the difference between WChPT and ChPT clear, we consider only
the case of the $N_f =2$ QCD with degenerate quark masses, and derive
the formulae for mass and decay constant of the pion as well as the axial 
Ward-Takahashi identity quark mass, as a function of the ``quark mass'' in 
the effective theory.
In section \ref{sec:analysis},
the derived formulae are applied to data of pion mass and the axial 
Ward-Takahashi identity quark mass calculated by the CP-PACS 
collaboration\cite{cppacs}. We show that data are consistent with
the formulae.
We have attempted the resummation of the most singular term, and
have derived the modified formulae in section~\ref{sec:resum}.
Our conclusions and discussions are given in section \ref{sec:conclusion}.

\section{Wilson Chiral Perturbation Theory}
\label{sec:wcpt}

\subsection{Derivation of effective Lagrangian}
It is difficult to derive the effective chiral Lagrangian for mesons
directly from lattice QCD with the Wilson-type quarks using the symmetry,
since the quark mass requires a counter term $m_c$, which diverges as $g^2/a$
near the continuum limit, so that $m_c a = O(1)$ and the conventional
power counting of $a$ fails.
Therefore, following the proposal\cite{SS,RS}, we overcome this problem
by first matching the lattice QCD to an effective continuum-like QCD including
the scaling violations into higher dimensional local operators\cite{Symanzik},
then match the latter to the effective Lagrangian for the 
Wilson chiral perturbation theory(WChPT).

Close to the continuum limit, the lattice QCD can be described by an effective
action in the continuum, which is expanded in power of $a$ as
\beqa
S_{\rm eff} &=& S_0 + a S_1 + a^2 S_2 + \cdots ,
\label{eq:qcd}
\eeqa
where $S_1$ contains chiral non-invariant terms only, while
$S_2$ contains chiral invariant as well as chiral non-invariant terms.
By using the equation of motion and the redefinition of the quark field,
quark mass and the coupling constant, only one term is relevant in $S_1$:
\beqa
S_1 &=& a r_1 \bar\psi \sigma_{\mu\nu} F_{\mu\nu} \psi + \cdots .
\eeqa
The similar analysis can be done for $S_2$\cite{SW}.

We now derive the effective Lagrangian of the WChPT from $S_{\rm eff}$, using 
the symmetry of $S_{\rm eff}$ such as parity, axis inter-change symmetry
(rotational invariance in the continuum limit), and the chiral symmetry.
The last one is explicitly broken not only by the quark mass $m$ but also
by the breaking terms in $S_1$ and $S_2$, whose coefficients are denoted as 
$r_i$( $i=1,2,3,\cdots $). One can make $S_{\rm eff}$ formally chiral invariant
by transforming $m$ and $r_i$'s to compensate the chiral variation of $\psi$ 
and $\bar\psi$. For example, if one writes the quark mass term as
\beqa
\bar\psi M P_R \psi + \bar \psi M^\dagger P_L \psi ,
\eeqa
this term is invariant under
\beqa
\psi&\rightarrow& \left(R P_R + L P_L\right)\psi , \qquad
\bar\psi\rightarrow \bar\psi\left(L^\dagger P_R + R^\dagger P_L\right) \\
M &\rightarrow& L M R^\dagger , \qquad
M^\dagger \rightarrow R M^\dagger L^\dagger ,
\eeqa
where $R$ and $L$ are SU($N_f$) chiral rotations.
The usual mass term is recovered by setting $M=M^\dagger = m$.
The similar transformations can be defined for $r_i$'s, but
we do not give them explicitly since the detail of them is irrelevant
for later discussion.
From this argument one concludes that the effective Lagrangian
of the WChPT should have this (generalized)
chiral $SU(N_f)_R\otimes SU(N_f)_L$ symmetry.

As mention in the introduction, we consider the $N_f=2$ case to make our 
argument simple and clear.  In this case, the chiral field for the 
pseudo-scalar mesons(pions) is given by
\beqa
\Sigma (x) &=& \Sigma_0 \exp \left\{ i \sum_{a=1}^3 \pi^a(x) t^a/f\right\}
= \Sigma_0\left[ \cos (\pi/f) + i \hat{\pi}^a t^a \sin (\pi/f)\right]
\eeqa
where $\pi^a (x)$ is the pion field, $t^a\equiv\sigma^a$ is the ordinary 
Pauli matrix and $f$ is the pion decay constant, whose experimental value is 
93 MeV.
The norm and the unit vector of the pion fields are given by
$\pi^2 = \pi\cdot\pi =\sum_a \pi^a \pi^a$, and $\hat\pi^a = \pi^a/\pi$,
respectively.
As discussed in ref.~\cite{SS}, 
the vacuum expectation value $\Sigma_0$ may have a complicated structure,
leading to the spontaneous breaking of parity-flavor symmetry, but
in this paper, we stay in the phase without this symmetry breaking,
so that $\Sigma_0 = {\bf 1}_{2\times2}$.
Under the chiral rotation, this field is transformed as
$ \Sigma \rightarrow  L  \Sigma R^\dagger$.
Under the transformation that $\pi \rightarrow -\pi$, called ``parity''
in this paper, $\Sigma \rightarrow \Sigma^\dagger$.

Using this field, we define the following naive operators
for Scalar(S), Pseudo-scalar(P), Vector(V), and Axial-vector(A):
\beqa
S^0 &=& \frac{1}{4}\tr \left(\Sigma + \Sigma^\dagger\right) = \cos (\pi/f),
\qquad
S^a = \frac{1}{4}\tr\ t^a\left(\Sigma + \Sigma^\dagger\right) = 0 \\
P^0 &=& \frac{1}{4}\tr \left(\Sigma - \Sigma^\dagger\right) = 0 ,\qquad
P^a = \frac{1}{4}\tr\ t^a\left(\Sigma - \Sigma^\dagger\right) = 
i\hat\pi^a\sin (\pi/f) \\
L_\mu^0&=&\frac{1}{2}\tr \left(\Sigma\partial_\mu \Sigma^\dagger\right)=0,
\qquad
L_\mu^a=\frac{1}{2}\tr\ t^a\left(\Sigma\partial_\mu \Sigma^\dagger\right)\\
R_\mu^0&=&\frac{1}{2}\tr \left(\Sigma^\dagger\partial_\mu \Sigma\right)=0,
\qquad
R_\mu^a =\frac{1}{2}\tr\ t^a\left(\Sigma^\dagger\partial_\mu \Sigma\right)\\
V_\mu^0 &=&\frac{1}{2}\left(L_\mu^0+R_\mu^0\right)=0,\qquad
A_\mu^0 =\frac{1}{2}\left(L_\mu^0-R_\mu^0\right)=0\\
V_\mu^a &=&\frac{1}{2} \left(L_\mu^a+R_\mu^a\right)
= i e^{abc} \hat\pi^b \sin (\pi/f)\partial_\mu [\hat\pi^c \sin (\pi/f) ] \\
A_\mu^a &=&\frac{1}{2} \left(L_\mu^a-R_\mu^a\right)
= i \left\{\hat\pi^a \sin (\pi/f)\partial_\mu [\cos (\pi/f)]
-\cos (\pi/f) \partial_\mu [\hat\pi^a \sin (\pi/f) ] \right\}
\eeqa
where the suffices $0$ and $a$ mean the flavor singlet and triplet,
respectively. We also introduce Left-handed(L) and Right-handed(R) currents
for later use.
Due to the speciality of the $N_f=2$ case, some of the
above operators are identically zero. Here we do not consider the Tensor(T)
operator, which must contain two derivatives, since it does not contribute
to the 1-loop calculation in this paper.

Now we construct the effective Lagrangian, which must be invariant under
parity, axis-interchange symmetry and the (generalized) chiral symmetry.
In the 1-loop calculation, which gives the main contribution at the 
next-to-leading order in the chiral perturbation theory, 
it is enough for us to  construct the effective Lagrangian up to
the order $m$, where $m$ is the quark mass in the effective theory.
On the other hand, we must include the $O(a^2)$ effect
to realize the massless pions at $a\not= 0$\cite{SS}.
At the next-to-leading order, $O(m^2)$ counter terms
(Gasser-Leutwyler coefficients) are also needed. We do not include, however, 
these terms in our effective Lagrangian, since we will not intend to
determine them in this paper. Instead we introduce arbitrary scale parameters
in $\log(m) $ terms which appear in 1-loop integrals.
Roughly speaking, we consider the situation that
$
1 \gg a \ge m\simeq  p^2 \ge a^2 \ge  m a \simeq p^2 a \ge m^2\simeq p^4
\simeq mp^2 $,
so that all terms up to $m a$ or $ p^2 a$ in this inequality
will be included in the effective Lagrangian.

The chirally invariant contribution at the leading order, which has the least
number of derivatives, is constructed from $L_\mu^a$ or $R_\mu^a$ as follows:
\beqa
2\sum_{a=1}^3 L_\mu^a L_\mu^a &=& 2\sum_{a=1}^3 R_\mu^a R_\mu^a = 
\tr\left[ \partial_\mu \Sigma^\dagger\partial_\mu\Sigma \right]\nn \\
&=&
2\left\{
\partial_\mu [\cos (\pi/f)] \partial_\mu [\cos (\pi/f)] 
+\partial_\mu [\hat\pi^a \sin (\pi/f)]\partial_\mu [\hat\pi^a \sin (\pi/f)]
\right\},\\
L_\mu^0 L_\mu^0 &=& R_\mu^0 R_\mu^0 = 0 .
\eeqa
Note that $R_\mu^a L_\mu^a$ term is prohibited by the parity invariance.
The chirally non-invariant parity-even term accompanied with
one power of $m$, $r_1 =O(a)$ or  $r_{i\ge 2} =O(a^2)$
is uniquely given by $S^0$.
The chirally non-invariant terms whose coefficients include 
$r_1^2 = O(a^2)$ or $r_1\cdot m = O(ma)$ are given by $(S^0)^2$, 
$ \sum_a (P^a)^2$ or $\tr (\Sigma + \Sigma^\dagger)^2$. 
For the $N_f=2$ case, however, the latter two terms are not independent, 
as evident from the expressions that 
$\sum_{a=0}^3 (P^a)^2 = (S^0)^2 -1$ and 
$\tr (\Sigma + \Sigma^\dagger)^2 \propto (S^0)^2$. 
An independent term at $O( a p^2)$ is given uniquely by
$  S_0\times \tr [ \partial_\mu \Sigma^\dagger\partial_\mu\Sigma ]$,
since $\tr [(\Sigma+\Sigma^\dagger) \partial_\mu \Sigma^\dagger
\partial_\mu\Sigma ]$ is not independent for SU(2).

Gathering all terms up to $m, p^2$, $a^2$ and $ m a, p^2 a$, 
the effective Lagrangian becomes
\beqa
L_{\rm eff}&=&\frac{f^2}{4}\left[1 + c_0 (S^0-1)\right]
\tr\left\{ \partial_\mu \Sigma^\dagger\partial_\mu\Sigma \right\}
- c_1 S^0 + c_2 (S^0)^2 ,
\eeqa
where parameters $c_0$, $c_1$ and $c_2$ have the  leading
$m$ and $a$ dependences as
\beqa
c_0 &=& W_0 a + O(m) \\
c_1 &=& W_1 a + B_1 m\\
c_2 &=& W_2 a^2 + V_2 m a + O(m^2) .
\eeqa
Since $c_0$ is dimensionless and $c_1$ and $c_2$ have the mass dimension 4,
$W_0 \sim \Lambda ( 1 + O(\Lambda a))$,
$W_1 \sim \Lambda^5 ( 1 + O(\Lambda a))$,
$W_2 \sim \Lambda^6 ( 1 + O(\Lambda a))$,
$V_2 \sim \Lambda^4 ( 1 + O(\Lambda a))$,
$B_1 \sim \Lambda^3 $,
where $\Lambda$ represents some mass scale of the theory such as 
$\Lambda_{\rm QCD}$.  
The (sub-leading) $a$ dependence of these parameters comes
from the chiral breaking terms of $a^2 S_2$ in the effective action
eq.(\ref{eq:qcd}), which correspond to $r_{i\ge 2} = O(a^2)$ terms
in $c_0$ and $c_1$, or $r_1\cdot r_{i\ge 2}= O(a^3)$ and 
$m\cdot r_{i\ge 2} = O(ma^2)$ terms in $c_2$.
Chirally invariant parameters such as $f$ receive $O(a^2)$ corrections from
chirally invariant $O(a^2)$ terms in $a^2 S_2$.
Note that $W_0$, $W_1$, $V_2 \sim O(a)$ if
non-perturbatively $O(a, ma)$ improved fermions
are employed for the lattice QCD action.

For later use, we define the operators in the effective theory,
which correspond to the ones in QCD up to $O(a)$:
\beqa
\left[ S^0 \right] &=& Z_S S^0 \{ 1 + c_S (S^0-1) \},\quad
\left[ P^a \right] = Z_P P^a \{ 1 + c_P (S^0-1) \} \\
\left[ V_\mu^0 \right] &=&  \tilde{c}_V \partial_\mu S^0,\quad
\left[ V_\mu^a \right] = Z_V V_\mu^a \{ 1 +  c_V(S^0-1) \} , \\
\left[ A_\mu^a \right] &=& Z_A\left\{ A_\mu^a \left( 1 + c_A (S^0-1)\right)
+\tilde{c}_A\partial_\mu P^a\right\} 
\eeqa
where $c_{S,P,V,A}$ and $\tilde{c}_{A,V}$ are $O(a)$ in general,
or $O(a^2)$ if the lattice action and operators are
non-perturbatively $O(a)$ and $O(ma)$ improved.

\subsection{Next-to-leading order calculations}
To perform the next-to-leading order (1-loop) calculation, we expand 
$L_{\rm eff}$  in terms of the pion field $\pi^a$ as
\beqa
L_{\rm eff}&=&\mbox{ const.} +\frac{1}{2}\left[
\partial_\mu \pi \cdot \partial_\mu \pi + \frac{c_1-2c_2}{f^2}\pi^2\right]
\nn \\
&+&\frac{1}{6f^2}\left[(\pi\cdot\partial_\mu\pi)^2-(1+\frac{3}{2}c_0)
(\partial_\mu\pi\cdot\partial_\mu\pi) \pi^2  \right]
+\frac{ (\pi^2)^2}{4! f^4}(8 c_2-c_1)
\eeqa
and the operators as
\beqa
\left[S^0\right]&=& Z_S (1-\frac{\pi^2}{2!f^2})\left(1-c_S\frac{\pi^2}{2!f^2}
\right)=Z_S \left[1-\frac{\pi^2}{2!f^2}(1+c_S)\right]\\
\left[P^a\right]&=&i Z_P \frac{\pi^a}{f}\left[1-\frac{\pi^2}{3!f^2}(1+3c_P)\right]\\
\left[V_\mu^a\right]&=&i Z_V  e^{abc}\frac{\pi^b\partial_\mu\pi^c}{f^2}
\left(1-\frac{\pi^2}{3!f^2}(1+3c_V)\right)\\
\left[V_\mu^0\right]&=&-\tilde{c}_V\frac{\pi\cdot\partial_\mu\pi}{f}
\left(1-\frac{\pi^2}{3!f^2} \right) \\
\left[A_\mu^a\right]&=& i Z_A \left[
(1+\tilde{c}_A)\frac{\partial_\mu\pi^a}{f}
-\frac{2\partial_\mu\pi^a \pi^2}{3f^3}(1+\frac{3c_A+\tilde{c}_A}{4})
+\frac{2 \pi^a \pi\cdot\partial_\mu\pi}{3 f^3}(1-\frac{\tilde{c}_A}{2})
\right] .
\eeqa

Using the pion propagator at the tree-level, which is given by
\beqa
\langle \pi^a(-p) \pi^b (p)\rangle_0 &=& \delta_{ab} \frac{1}{ p^2 + m_0^2} \\
m_0^2 &=& \frac{c_1-2 c_2}{f^2},
\eeqa
we evaluate loop integrals as usual
\beqa
\langle \pi^a (x) \pi^b (x) \rangle &=& \delta_{ab} I
=\delta_{ab} \frac{m_0^2}{16\pi^2}\log \frac{m_0^2}{\Lambda^2}
\label{eq:int1}\\
\langle \partial_\mu \pi^a (x) \partial_\nu \pi^b (x) \rangle &=& 
\delta_{ab} \frac{\delta_{\mu\nu}}{4} \left(- m_0^2 I\right),
\label{eq:int2}
\eeqa
where we introduce an arbitrary scale parameter $\Lambda$ resulting after
removals of power  divergences of loop integrals by the local counter terms.
Therefore, although we use the same symbol,
this $\Lambda$ varies depending on physical observables.

The inverse pion propagator at the 1-loop level is calculated as
\beqa
L_{\rm eff}^{(2)} &=&
\frac{1}{2}(\partial_\mu\pi)^2\left\{1-\frac{I}{3f^2}
(2+\frac{9 c_0}{2})\right\}
+\frac{1}{2}\pi^2\left\{ m_0^2\left(1-\frac{I}{6f^2}(1-9c_0)\right)
+\frac{5c_2 I}{f^4} 
\right\} \nn \\
&=&\frac{1}{2}\left[ (\partial_\mu \pi_R )^2 + m_\pi^2 \pi_R^2\right]
\eeqa
where
\beqa
\pi &=& Z^{1/2} \pi_R \\
Z &=& \left[1-\frac{I}{3f^2}(2+\frac{9c_0}{2})\right]^{-1} \\
m_\pi^2 &=& m_0^2\left[ 1+\frac{m_0^2}{32\pi^2 f^2}(1+6 c_0)
\log \frac{m_0^2}{\Lambda^2} + \frac{5 c_2}{16\pi^2f^4}
\log \frac{m_0^2}{\Lambda^2} \right] .
\eeqa

For the axial-vector current, we obtain
\beqa
\langle [ A_\mu^a] (x) \pi_R^b (y)\rangle &=&\delta_{ab}
\frac{i Z_A}{f} \langle \partial_\mu\pi_R^a(x)\pi_R^b(y)\rangle_0 Z^{1/2}
\left[(1+\tilde{c}_A)-\frac{I}{3f^2}
\left(4+\frac{9c_A-3\tilde{c}_A}{2}\right)
\right],
\eeqa
therefore the decay constant at the 1-loop order becomes
\beqa
f_\pi &=& \frac{i Z_A }{\sqrt{2} f^2}
f(1+ \tilde{c}_A)\left[1
-\frac{m_0^2}{16\pi^2f^2}\left(1+\frac{3c_A}{2}-\frac{11\tilde{c}_A}{6}
-\frac{3c_0}{4}
\right)\log \frac{m_0^2}{\Lambda^2}\right] .
\eeqa
Taking $Z_A =- i\sqrt{2} f^2$, we have
\beqa
f_\pi &=& f(1+\tilde{c}_A) \left[1
-\frac{m_0^2}{16\pi^2f^2}\left(1+c_{f_\pi}\right)
\log \frac{m_0^2}{\Lambda^2}\right]
\eeqa
where $c_{f_\pi} =3c_A/2-11\tilde{c}_A/6-3 c_0/4$.
Note that $f_\pi$ receives an $O(a)$ correction even in the chiral limit:
$f_\pi = f (1+\tilde{c}_A) $.

Similarly, we have
\beqa
\langle \partial_\mu [A_\mu^a] (x) \pi_R^b(y) \rangle &=&
\langle \pi_R^a(x) \pi_R^b(y)\rangle_0 \sqrt{2} f m_\pi^2 Z^{1/2}
\left[(1+\tilde{c}_A)-\frac{I}{3f^2}
\left(4+\frac{9c_A-3\tilde{c}_A}{2}\right)
\right]\\
\langle [P^a](x)\pi_R^b (y)\rangle &=&
i \frac{Z_P}{f}\langle \pi_R^a(x)\pi_R^b(y)\rangle Z^{1/2}
\left[1-\frac{5I}{3!f^2}(1+3c_P)\right].
\eeqa
Then the PCAC quark mass $m_{\rm AWI}$ is given by
\beqa
m_{\rm AWI} &=& \frac{\langle \partial_\mu [A_\mu^a] (x) \pi_R^b(y) \rangle} 
{\langle [ P^a] (x) \pi_R^b (y)\rangle}
\nn\\
&=& \frac{\sqrt{2} f^2}{iZ_P} m_\pi^2 (1+\tilde{c}_A)
\left[ 1 - \frac{m_0^2}{32\pi^2 f^2}(1+ 3c_A-11\tilde{c}_A/3-5c_P)
\log \frac{m_0^2}{\Lambda^2}\right] \nn\\
&=& \frac{1+\tilde{c}_A}{2B_0} m_0^2 
\left[ 1 + \frac{m_0^2 c_{m_{\rm AWI}}+ 10 c_2/f^2}{32\pi^2 f^2}
\log \frac{m_0^2}{\Lambda^2}\right] ,
\eeqa
where $1/(2B_0) = \sqrt{2} f^2/(iZ_P)$ and
$c_{m_{\rm AWI}}= 6c_0 - 3c_A+11\tilde{c}_A/3+5c_P$.

Let us recall the leading $m$ and $a$ dependences of the parameters:
\beqa
c_0 &=& W_0 a, \quad 
c_1 = W_1 a + B_1 m , \quad
c_2 = W_2 a^2 + V_2 ma \\
c_P &=& W_P a, \quad
c_A = W_A a , \quad
\tilde {c}_A = \tilde{W}_A a ,
\eeqa
and then the pion mass at tree level is written as
\beqa
m_0^2 &=& \frac{c_1-2 c_2}{f^2}
=\frac{ m(B_1 -2V_2 a)+ aW_1 -2 a^2 W_2}{f^2}
= A ( m - m_c) \equiv A m_R
\eeqa
where
\beqa
A&=& \frac{B_1 -2 a V_2}{f^2},\qquad
m_c = -a\frac{W_1 -2 a W_2}{B_1-2a V_2},\qquad
m_R = m - m_c .
\eeqa
Here it is noted that $m_c=O(a)$ does not correspond to $1/(2K_c)$ in 
lattice QCD,
since the $1/a$ contribution to the quark mass is already subtracted in $m$.
Furthermore, for $m < m_c$, pion would become tachyonic ($m_0^2 < 0$).
As discussed in ref.~\cite{SS}, however, as long as $c_2=W_2 a^2+V_2 m_c a =
O(a^2) > 0$, the parity-flavor symmetry  breaking phase 
transition\cite{aoki1,aoki2,aoki3} occurs at $m=m_c=O(a)$, so that 
$m_0^2$ is always positive.
In other words, the $O(a^2)$ contribution in $c_2$ is necessary for the 
consistency between the PCAC relation ($m_\pi^2 \sim m_q$) and the absence of 
tachyons\footnote{On the other hand, if $c_2 < 0$, no massless pion appears
\cite{SS}.}.

We summarize the result of the 1-loop calculation in terms of $m_R$ and $a$:
\beqa
m_\pi ^2 &=& A m_R \left[
1+\frac{m_R( A + w_1 a) }{32\pi^2f^2} \log \frac{ A m_R}{\Lambda^2} 
+ \frac{w_0 a^2 }{32\pi^2f^2} \log \frac{ A m_R}{\Lambda_0^2} \right] \\
m_{\rm AWI} & = & A_0 m_R \left[
1+\frac{m_R w_1^{\rm AWI} a }{32\pi^2f^2} \log \frac{ A m_R}{\Lambda_{\rm
AWI}^2} 
+ \frac{w_0 a^2 }{32\pi^2f^2} \log \frac{ A m_R}{\Lambda_0^2} 
\right] \\
f_\pi &=& f(1+\tilde{c}_A) \left[
1-\frac{m_R(A+w_1^{\rm decay}a)}{16\pi^2f^2}\log 
\frac{ A m_R}{\Lambda_{\rm decay}^2} 
\right] 
\eeqa
where
\beqa
w_1 &=& 6 W_0 +\frac{10 V_2}{f^2} , \quad
w_0 = \frac{10}{f^2}\left( \frac{m_c V_2}{a}+W_2\right) \\
w_1^{\rm AWI} &=& w_1 - 3 W_A+\frac{11}{3}\tilde{W}_A+5 W_P , \quad
w_1^{\rm decay} =\frac{3}{2}W_A-\frac{11}{6}\tilde{W}_A-\frac{3}{4} W_0 \\
A_0 &=&\frac{A(1+\tilde{c}_A)}{2B_0} \simeq 1+O(a) .
\eeqa
Note that here $m_c/a = O(1)$ and we recover the distinction among
scale parameters ($\Lambda$, $\Lambda_0$, $\Lambda_{\rm AWI}$
or $\Lambda_{\rm decay}$).

These results reveal the following features of the WChPT.
In general the chiral log terms($m_R\log m_R$) receive $O(a)$ 
scaling violation.
In addition to this, the $a^2$ contribution generates
$\log m_R$ term in $m_\pi^2$,
which is more singular as a function of $m_R$ than the usual
chiral log term, $m_R\log m_R$.
Furthermore, both $m_R\log m_R$ and $\log m_R$ terms are generated in 
$m_{\rm AWI}$ by the scaling violations, 
$O(a)$ for the former and $O(a^2)$ for the latter.
The coefficient of $\log m_R$ term in $m_{\rm AWI}$ is same as the one
in $m_\pi^2$.

In the next section we employ the above formulae to fit the full QCD data
obtained by the CP-PACS collaboration\cite{cppacs}.

\section{Analysis of CP-PACS data}
\label{sec:analysis}
In this section, we apply the WChPT formulae to $m_\pi^2$ and $m_{\rm AWI}$
in the $N_f=2$ full QCD with the clover quark action\cite{cppacs}.

\subsection{Data sets and WChPT formulae}
The CP-PACS collaboration has performed the large scale full QCD simulations
with the RG improved gauge action and $N_f=2$ (tadpole improved) clover quark 
action, at 4 different lattice spacings $a$ and 4 different quark masses at
each $a$, as summarized in table~\ref{tab:cppacs}.
In ref.~\cite{cppacs} the data for $m_\pi^2$ and $m_{\rm AWI}$ have been 
published.
Unfortunately the data for $f_\pi$ at each quark mass are not available.

We define the quark mass $m_R$ in the WChPT theory in terms of the
hopping parameter $K$ in lattice QCD as
\beqa
m_R &=& Z_m ( 1 + b_m a \frac{m}{u_0}) \frac{m}{u_0},
\quad m a =\frac{1}{2K}-\frac{1}{2K_c},
\eeqa
where $K_c$ is the critical hopping parameter, and
$u_0$ is the tadpole improvement factor, given by
$u_0 = \displaystyle \left(1-\frac{0.8412}{\beta}\right)^{1/4}$.
This $m_R$ is identical to the renormalized VWI quark mass 
in ref.~\cite{cppacs}.
By definition, $m_\pi^2 =0$ at $m_R=0$ in lattice QCD.
We identify this $m_R$ in lattice QCD with $m_R$ in the WChPT,
since $m_0^2$, and therefore $m_\pi^2$,  must vanish at $m_R=0$
in the WChPT. 
We also use the renormalized $m_{\rm AWI}$ defined as
\beqa
m_{\rm AWI} &=& \frac{Z_A}{Z_P}m_{\rm AWI}^{\rm bare} .
\eeqa

We employ the following fitting forms for $m_\pi^2$ and $m_{AWI}$
\beqa
m_\pi^2 &=& A m_R \left[ 1 + \frac{m_R A+m_R a w_1}{32\pi^2 f^2}
\log \left(\frac{A m_R}{\Lambda^2}\right)
+\frac{a^2 w_0}{32\pi^2 f^2}\log \left(\frac{A m_R}{\Lambda_0^2}\right)
 \right]  
\label{eq:fit-mpi} \\
m_{\rm AWI}&=& A_0 m_R \left[ 1 + \frac{m_R a w_1^{\rm AWI}
}{32\pi^2 f^2}\log  \left(\frac{A m_R}{\Lambda_{\rm AWI}^2}\right)
+ \frac{a^2 w_0}{32\pi^2 f^2}\log  \left(\frac{A m_R}
{\Lambda_0^2}\right)
\right] .
\label{eq:fit-mawi}
\eeqa

\subsection{Results}

We first fit the data at each $a$ separately.
Since there are only 4 data per observables at each $a$, it is impossible
to fit an individual observable, $m_\pi^2$ or $m_{\rm AWI}$,
as a function of $m_R$ using eq.(\ref{eq:fit-mpi}) or eq.(\ref{eq:fit-mawi}),
each of which contains 4 or more parameters.
Therefore, we try to fit $m_\pi^2$ and $m_{\rm AWI}$ simultaneously.
Since $f$ can not be determined without data of $f_\pi$, we fix
$f = 93$ MeV\footnote{We have also performed the fit using measured values of 
$f_\pi$ in the chiral limit at each $\beta$\cite{cppacs}. We have found that
qualities of the two fits are similar.}.
Even in the simultaneous fit, 
the number of independent fitting parameters is still too large.
Since theoretically $A_0=1$ in the continuum limit
and the fit with $A_0=1$ becomes more stable, we fix $A_0=1$ in our fit.
In order to reduce a number of parameters further,
we set $\Lambda_{\rm AWI}=\Lambda_0=\Lambda$,
so we finally have 6 independent parameters,
$K_c$, $A$, $\Lambda$, $w_1$, $w_1^{\rm AWI}$ and $w_0$, for 8 data points.

\begin{figure}[htb]

\centering{
\hskip -0.0cm
\includegraphics[width=150mm,angle=0]{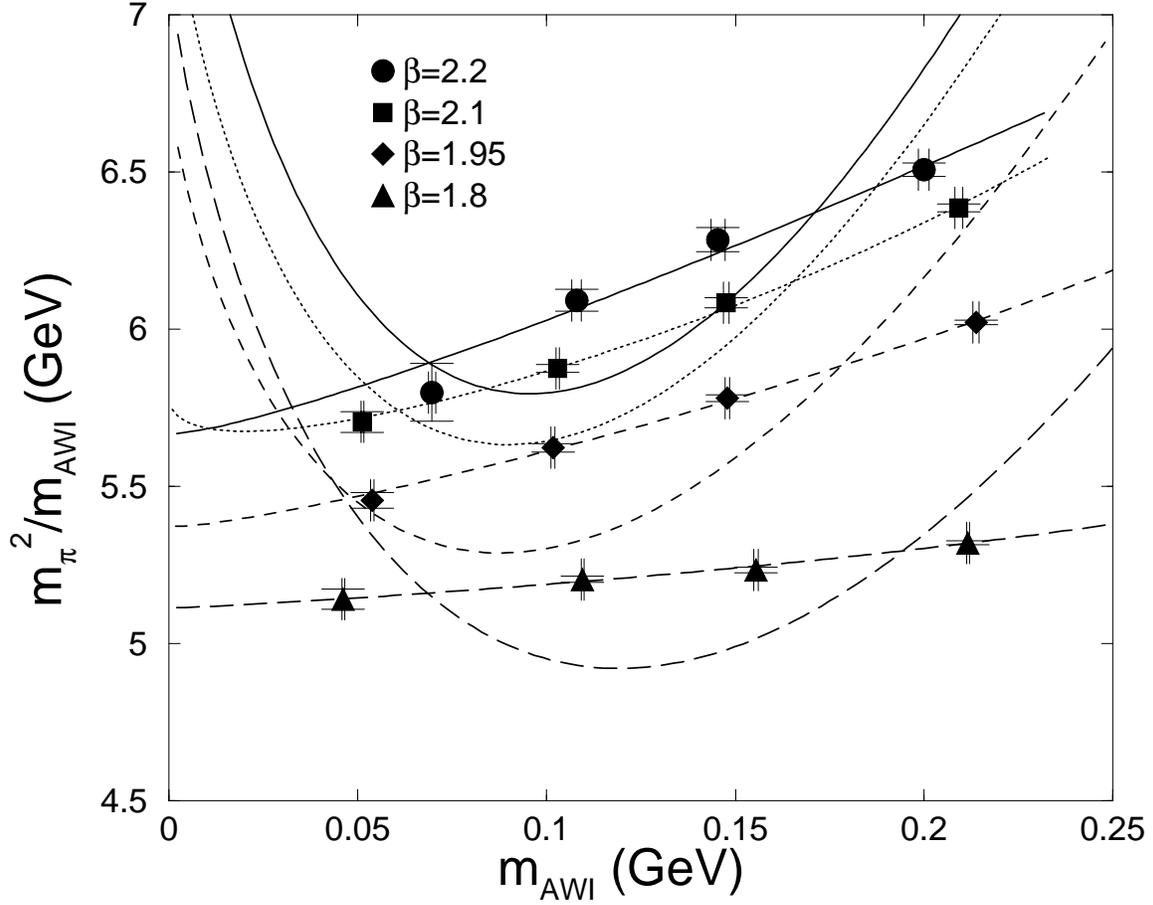}     
}
\caption{The WChPT fits for $m_\pi^2$ and $m_{\rm AWI}$ at each $\beta$.
Results are shown for $m_\pi^2/m_{\rm AWI}$ as a function of $m_{\rm AWI}$.
For comparison the standard ChPT fits ($w_1=w_0=0$) are also included.}
\label{fig:WChPT}
\end{figure}

Fig.~\ref{fig:WChPT} shows data and fits for $m_\pi^2/m_{\rm AWI}$  
as a function of $m_{\rm AWI}$ at each $a$. 
For comparison,
the results by the fit with the standard chiral perturbation theory
($w_1=w_0=0$) are also given. 
It is manifest that the WChPT fits perform much better than
the ChPT fits.
The parameters extracted from the fits are given in table~\ref{tab:WChPT}.
Note however that $\chi^2$/dof  shown in the table has not been reliably 
estimated due to the correlation between $m_\pi^2$ and $m_{\rm AWI}$,
which is not given in ref.~\cite{cppacs}.

\begin{figure}[htb]

\centering{
\hskip -0.0cm
\includegraphics[width=80mm,angle=0]{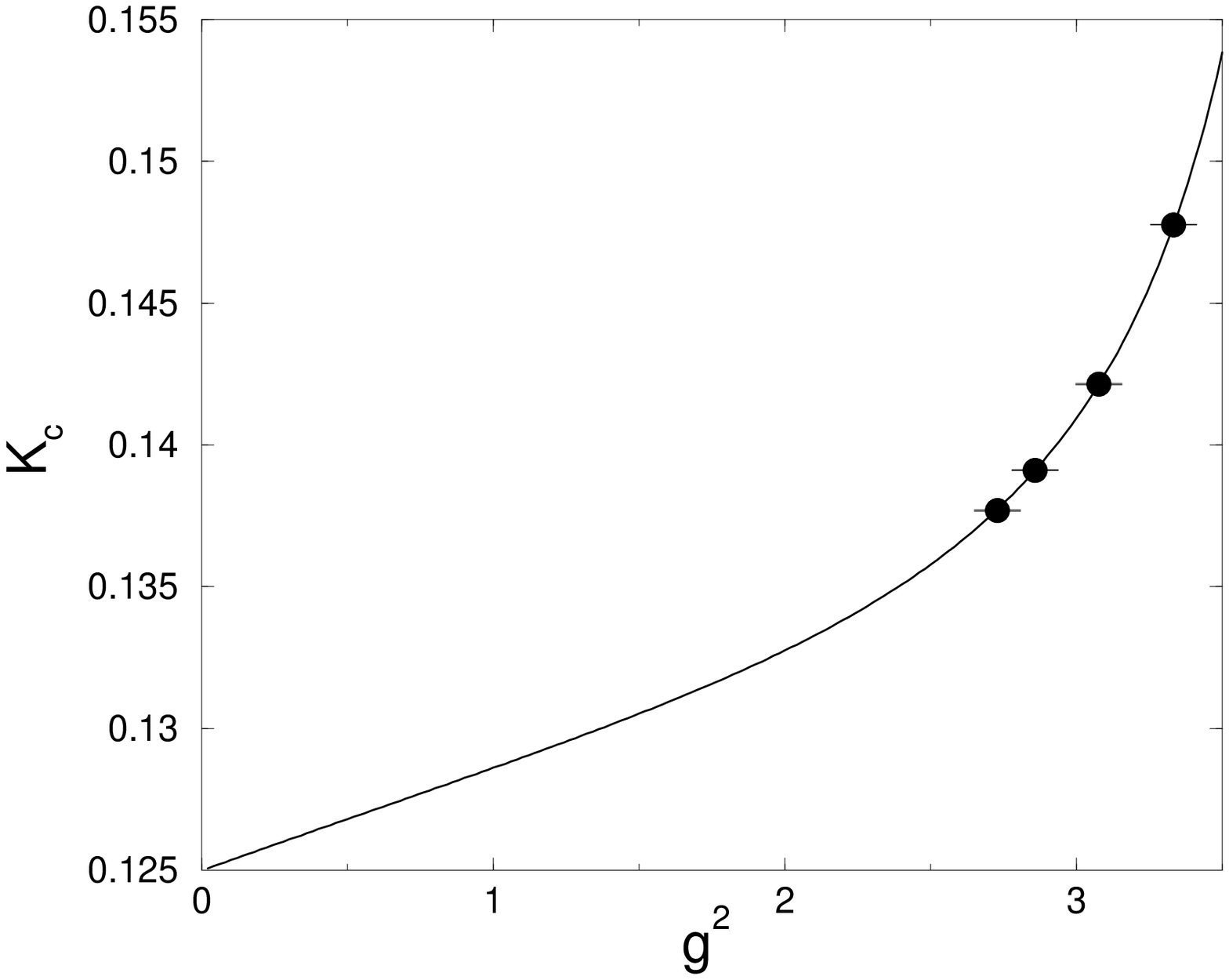}     
\includegraphics[width=80mm,angle=0]{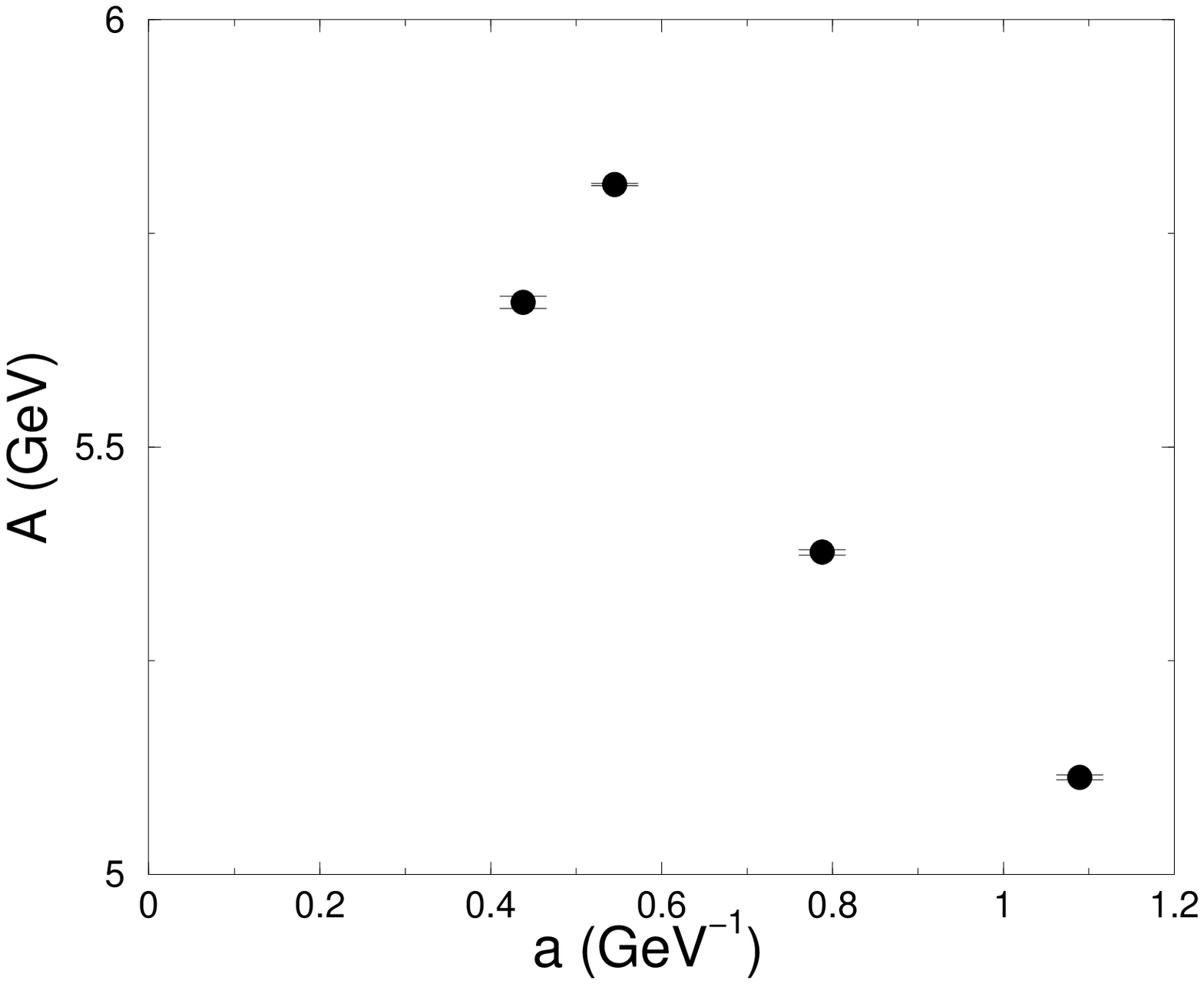}     
}
\centering{
\hskip -0.0cm
\includegraphics[width=80mm,angle=0]{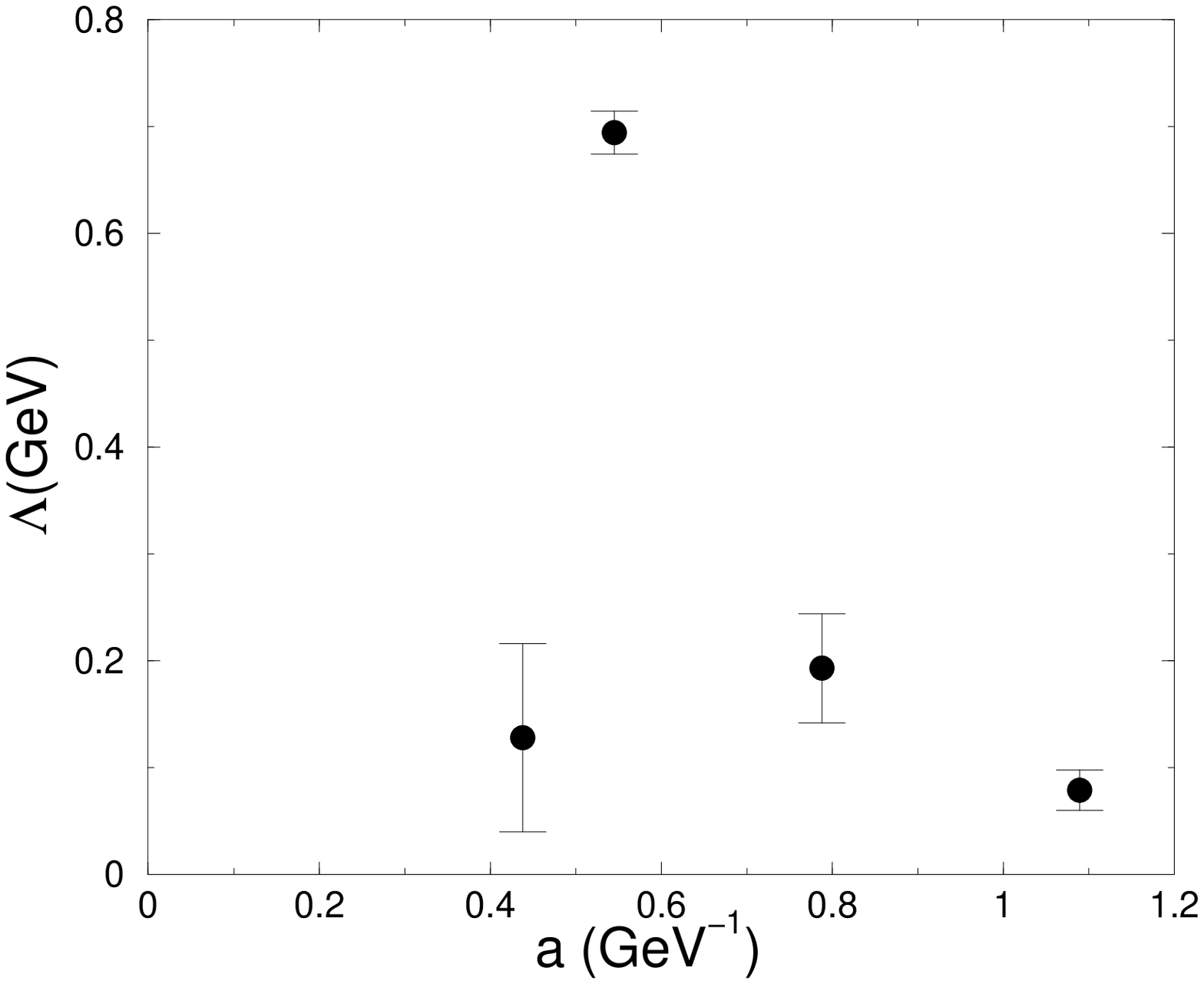}     
\includegraphics[width=80mm,angle=0]{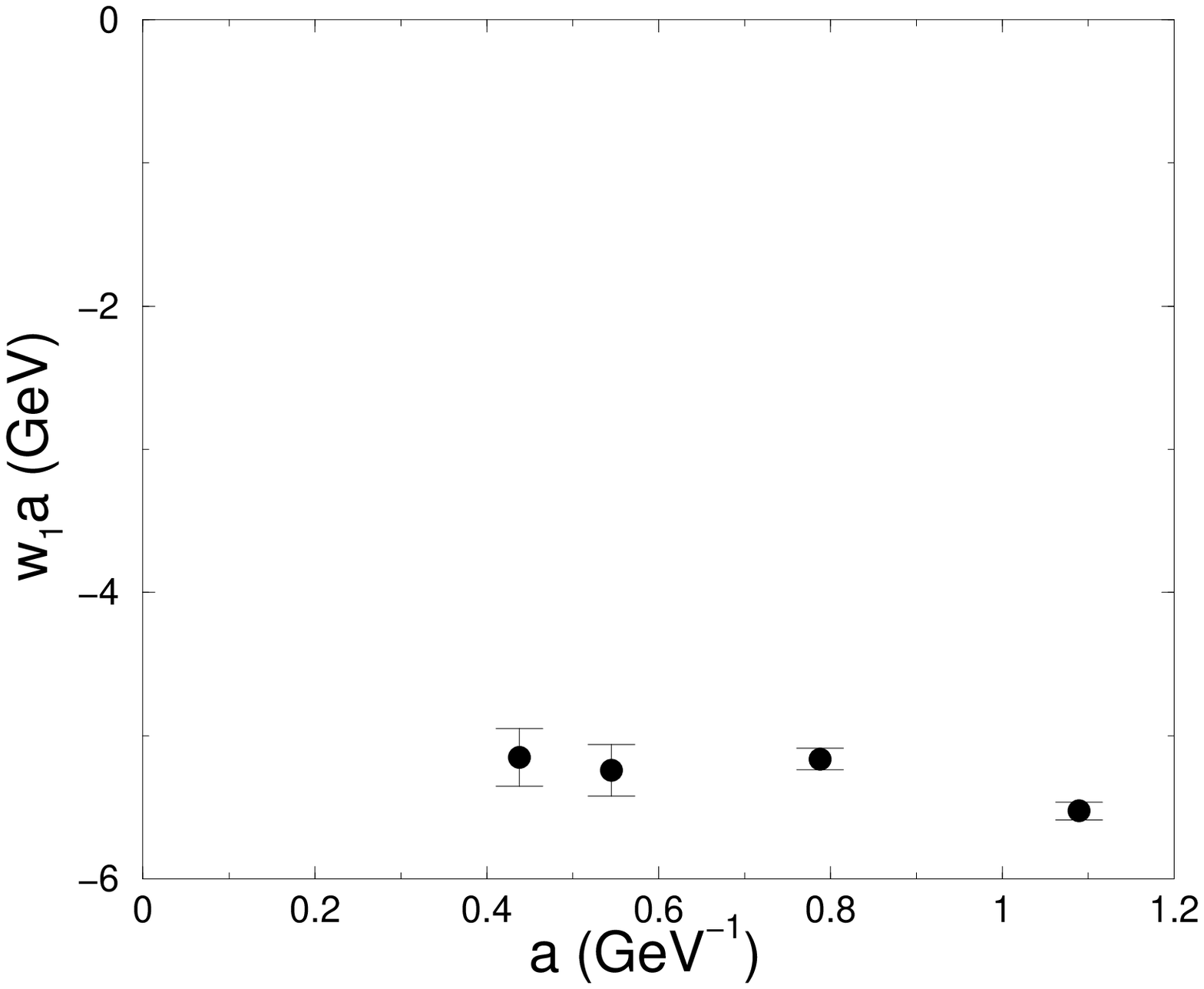}     
}
\centering{
\hskip -0.0cm
\includegraphics[width=80mm,angle=0]{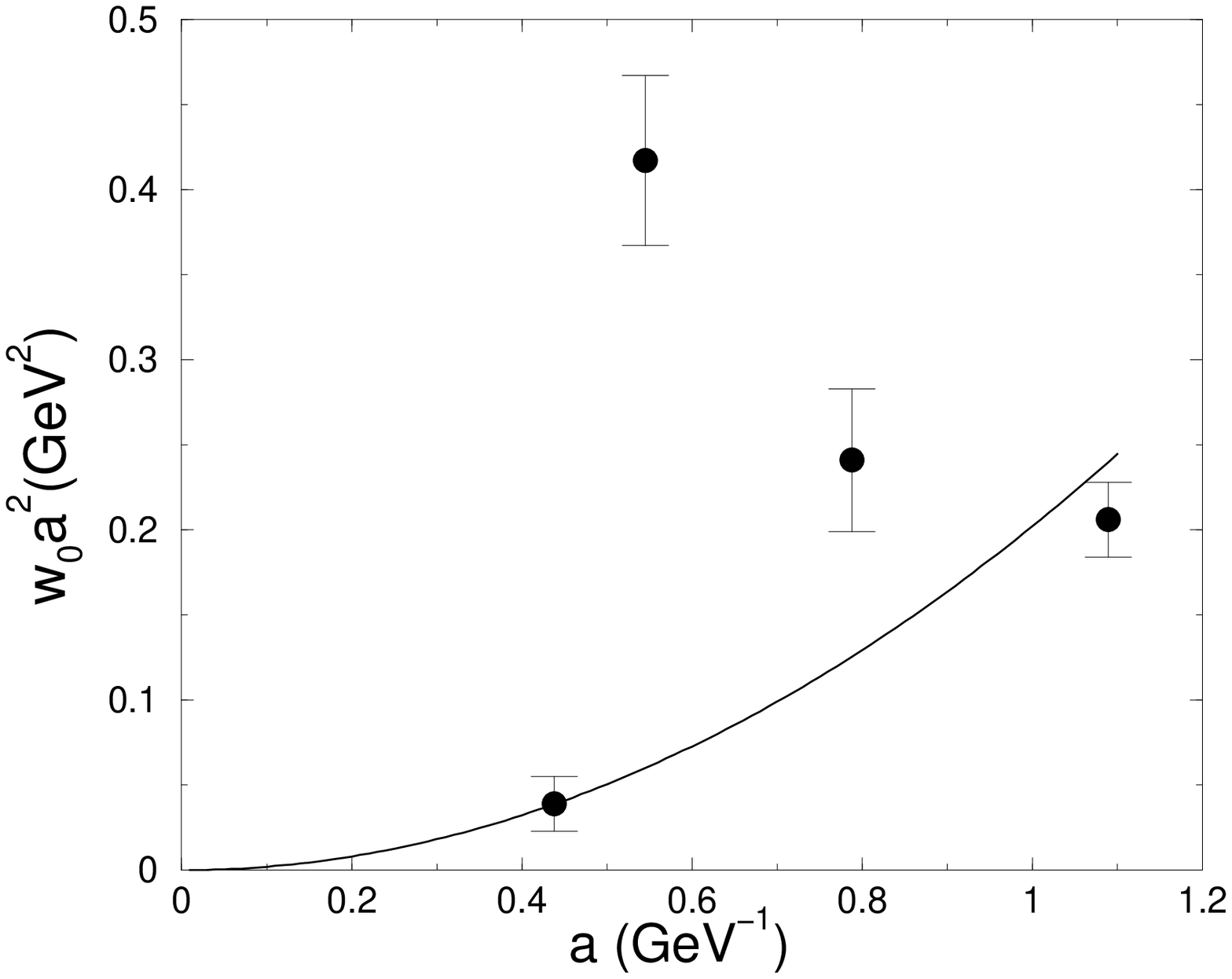}     
\includegraphics[width=80mm,angle=0]{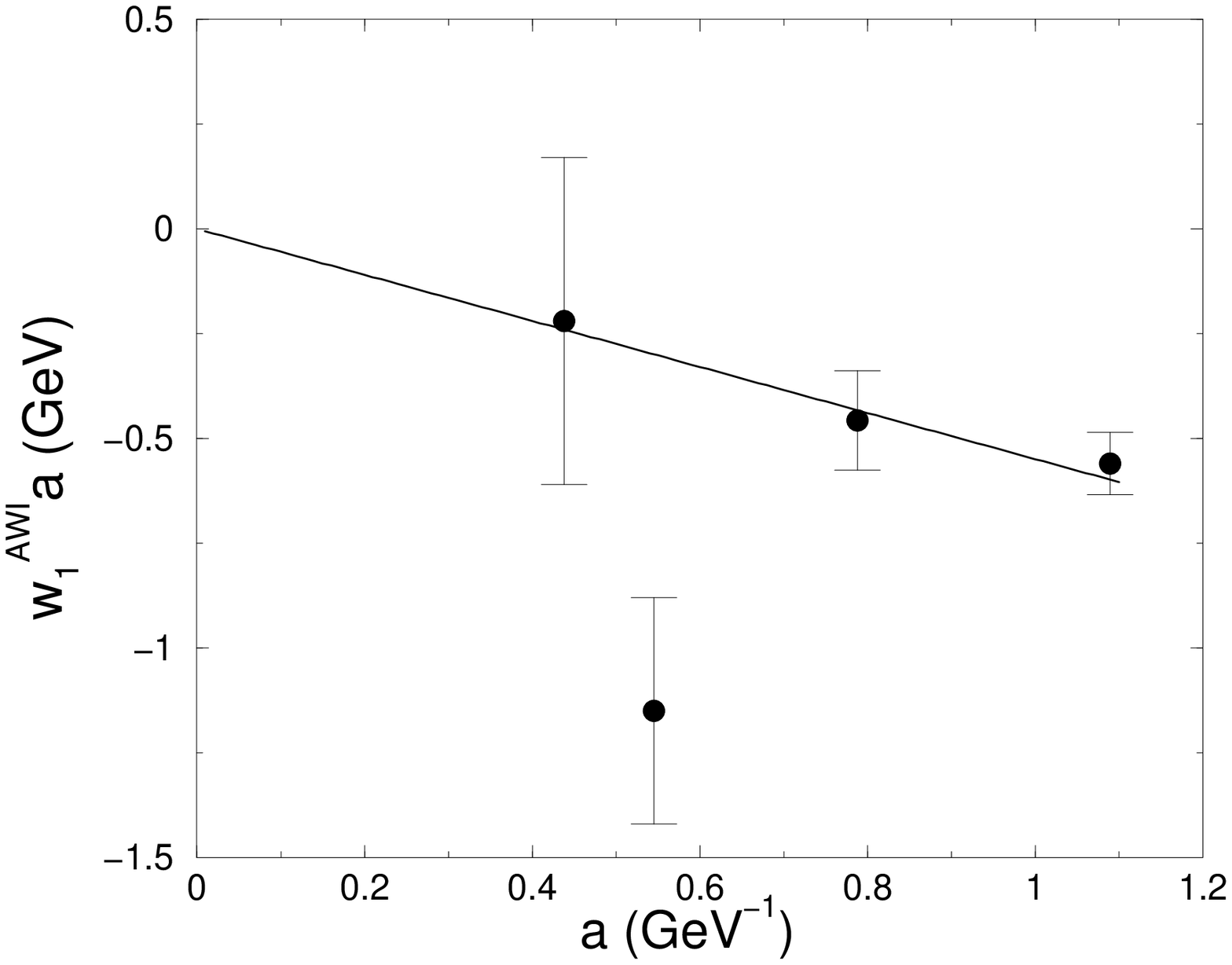}     
}
\caption{The fit parameters as a function of $a$ or $g^2$.}
\label{fig:param-a}
\end{figure}

In Fig.~\ref{fig:param-a}, $A$, $\Lambda $, $w_1 a$, $w_1^{\rm AWI} a$ and $w_0 a^2$ 
are plotted as a function of $a$, together with $K_c$ as a function of 
the bare gauge coupling constant $g^2$.
While $A$, $\Lambda$ and $w_1 a$ are too scattered to be fitted, 
$K_c$, $w_0 a^2$ and $ w_1^{\rm AWI} a$ may be fitted as 
\beqa
K_c &=& \frac{1}{8}\cdot\frac{1+d_0(K_c) g^2 + d_1(K_c) g^4+d_2(K_c) g^6}
{1+(d_0(K_c)-0.02945)g^2}
\eeqa
where 0.02945 is the 1-loop coefficient\cite{ANTU} and
\beqa
w_1^{\rm AWI} a &=& d_0(w_1) a, \qquad w_0 a^2= d_0(w_0) a^2 .
\eeqa
Fit curves are also shown in Fig.~\ref{fig:param-a}, and the extracted 
parameters are given in the column (a) of table~\ref{tab:global}.

\begin{figure}[htb]

\centering{
\hskip -0.0cm
\includegraphics[width=130mm,angle=0]{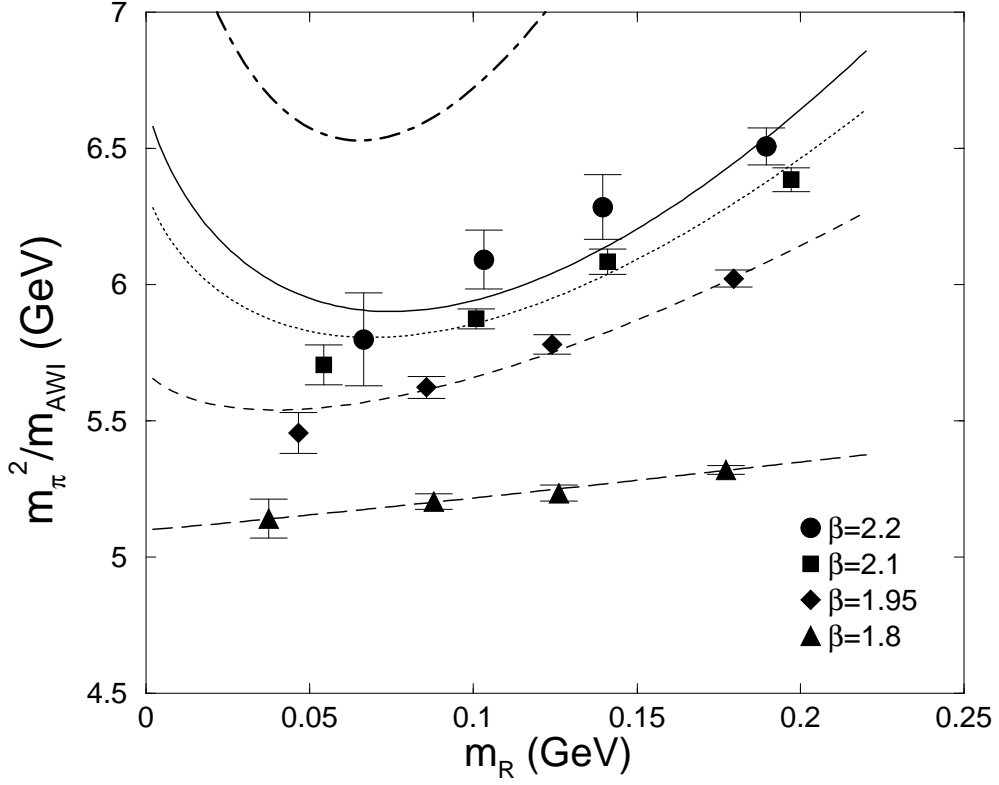}     
}
\caption{The WChPT fits for $m_\pi^2/m_{\rm AWI}$ as a function
of $m_R$ and $a$. Results are shown for $m_\pi^2/m_{\rm AWI}$ as a function of $m_{R}$.
}
\label{fig:pi.awi-mr}
\vspace{5mm}
\end{figure}

To determine $a$ dependences of $A$, $\Lambda$ and $w_1 a$, we have
fitted $m_\pi^2/m_{\rm AWI}$ as a function of both $m_R$ and $a$, using the following 
formula derived from eqs.(\ref{eq:fit-mpi},\ref{eq:fit-mawi}) with
$\Lambda_{\rm AWI}=\Lambda$:
\beqa
\frac{m_\pi^2}{m_{\rm AWI}} &=&
\frac{A}{A_0}\left[
1+\frac{(A+ \Delta w_1 a)m_R}{32\pi^2f^2}\log\left(\frac{Am_R}{\Lambda^2}
\right)\right]
\label{eq:ratio}
\eeqa
where 
\beqa
A&=& d_0(A)\left( 1+ d_1(A) a+ d_2(A) a^2\right),\qquad
A_0 = 1+ d_0(A_0) a \\
\Lambda &=& d_0(\Lambda)\left( 1+ d_1(\Lambda) a^2\right), \qquad
\Delta w_1= w_1-w_1^{\rm AWI} = d_0(\Delta w_1) a .
\eeqa
No $\log m_R$ term is presented in eq.(\ref{eq:ratio}).
Note however that $\log m_{\rm AWI}$ term appears again if
we replace $m_R$ in the right-hand side of eq.(\ref{eq:ratio})
with $m_{\rm AWI}$, due to the presence of the $\log m_R$ term in 
eq.(\ref{eq:fit-mawi}).
With $K_c$ fixed to  the values in table~\ref{tab:WChPT},
the fit works well, as shown in Fig.~\ref{fig:pi.awi-mr}, and the
fitted parameters are given in the column (b) of table~\ref{tab:global}.

We roughly estimate the size of each parameter, $B_1$, $V_2$, $W_{1,2,3}$ 
from the continuum extrapolations of $A$, $w_1$, $w_0$ and $m_c$.
Since we can not separate the $1/a$ contribution in $1/K_c$, however,
$m_c$ can not be extracted. Therefore, we simply set $m_c=0$, giving
that $W_1 = 2 a W_2$; the leading contribution of $W_1$ vanishes.
To reduce the number of the parameters further, we set $W_0=0$.
Then extracting $B_1$, $W_2$ and $V_2$ as
\beqa
B_1 &=& f^2 d_0(A) \equiv \left(\Lambda_{B_1}\right)^3\\
W_2 &=& \frac{f^2d_0(w_0)}{10} \equiv \left(\Lambda_{W_2}\right)^6\\
V_2&=&\frac{f^2 d_0(w_1)}{10}=\frac{f^2 (d_0(w_1^{\rm AWI})+d_0(\Delta w_1))}{10}
\equiv -\left(\Lambda_{V_2}\right)^4,
\eeqa
we obtain $\Lambda_{B_1}= 0.41$ GeV, $\Lambda_{W_2}= 0.24$ GeV and
$\Lambda_{V_2}= 0.21$ GeV.  
These $\Lambda_X$ takes a reasonable value, $\Lambda_X = 0.2\sim 0.4$ GeV.
If $ a \Lambda_X > m/\Lambda_X$, $O(a)$ terms become more important 
than $m_R$ terms. With $\Lambda_X= 0.2\sim 0.4$ GeV,
this condition at $a^{-1}$= 1 GeV or $a^{-1}$= 2 GeV
corresponds to $m_R < 40\sim 160$ MeV or $m_R < 20\sim 80$ MeV,
respectively.

\subsection{Validity of the (W)ChPT}

\begin{figure}[htb]

\centering{
\hskip -0.0cm
\includegraphics[width=80mm,angle=0]{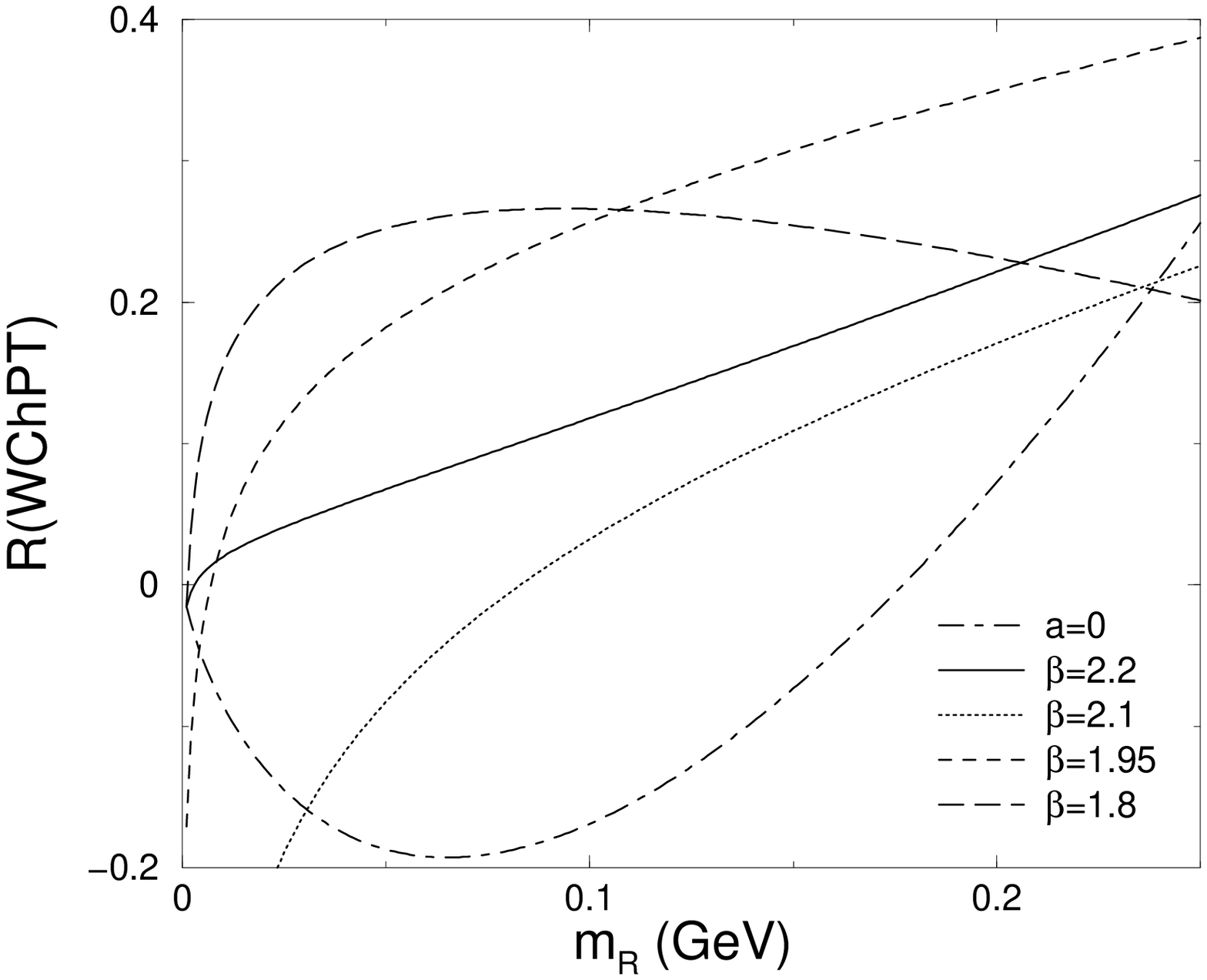}     
\includegraphics[width=80mm,angle=0]{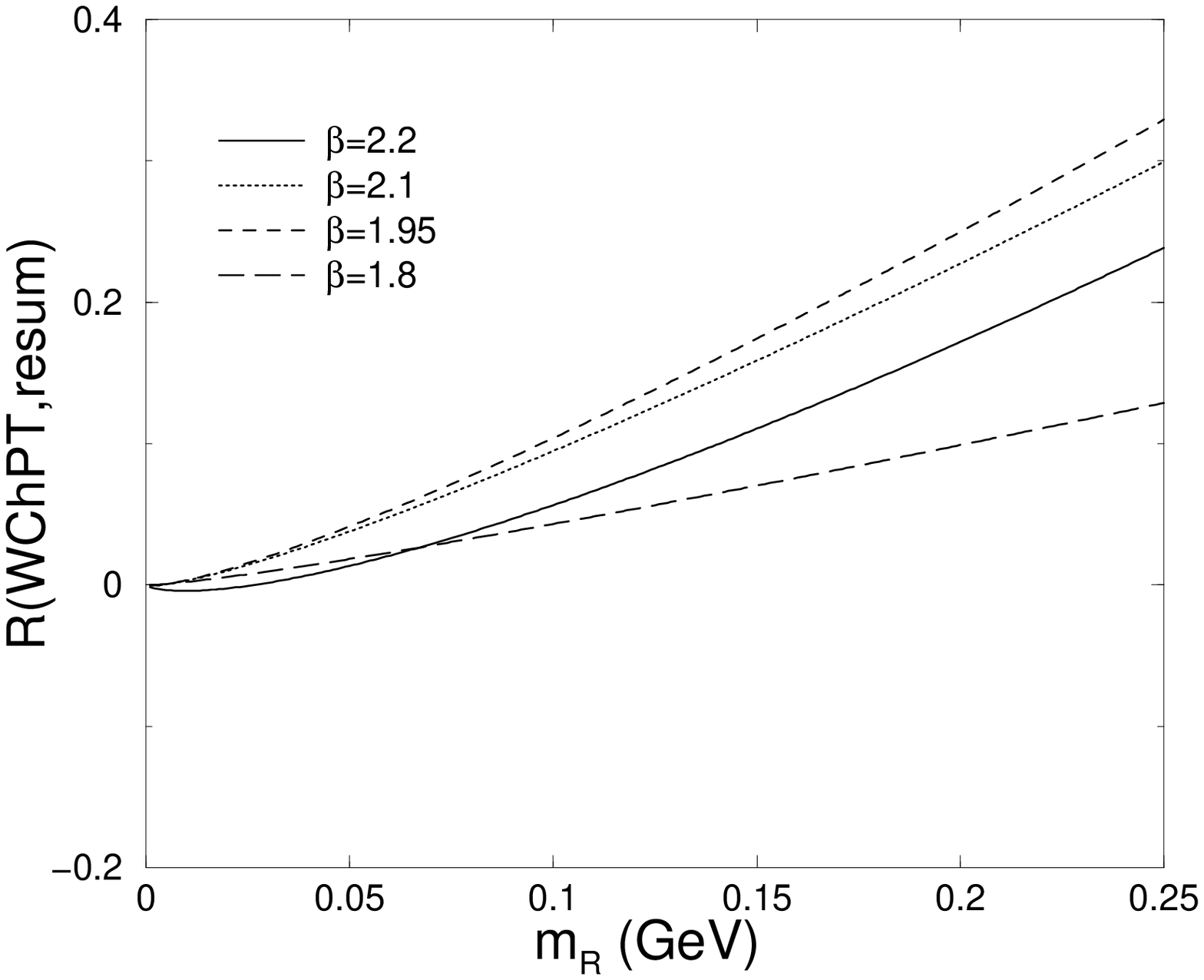}     
}
\caption{{\bf Left:} The relative size of the next-to-leading contribution to the
leading one in the WChPT as a function of the quark mass $m_R$ 
at $\beta$=1.8,1.95,2.1 and 2.2, together with the one in the continuum limit
(ChPT). {\bf Right:} Same quantities in the resummed WChPT.}
\label{fig:next}
\end{figure}

We now estimate the relative size size of the next-to-leading contribution 
to the leading contribution in the WChPT for $m_\pi^2$: 
\beqa
R(\mbox{WChPT})&=&
\frac{m_R(A+a w_1)+a^2 w_0}{32\pi^2 f^2}\log\left(\frac{A m_R}
{\Lambda^2}\right) 
\eeqa
for the WChPT at finite $a$, 
where parameters $A$, $\Lambda$, $w_1$ and $w_0$ depend on $a$.
We plot $R$(WChPT) in Fig.~\ref{fig:next} at $a$(GeV$^{-1}$)
=0, 0.44($\beta=2.2$), 0.55($\beta=2.1$), 0.79($\beta=1.95$) and
1.1($\beta=1.8$).
While the 1-loop contribution takes reasonable values, 10\% $\sim$ 30 \% ,
at 0.1 GeV $ < m_R < $  0.2 GeV for all $a$,
the contribution from $\log m_R$ in the WChPT diverges as 
$m_R\rightarrow 0$. This might invalidate the WChPT in the chiral limit.
We will consider this problem in the next section.

\section{Resummation of $\log m_R$ terms}
\label{sec:resum}
As evident from the analysis in the previous subsection,
$\log m_R$ contribution becomes larger and larger toward the chiral limit,
so that we can not neglect ``higher order'' term such as $(\log m_R)^n$
($n=2,3,\cdots$). We must perform a resummation of $\log m_R$ term
at all orders. 
Since it is possible in principle but difficult in practice to calculate 
$(\log m_R)^n$ contribution at $n$-loop order,
we derive resummed formulae from a different point of view.

As discussed in refs.~\cite{aoki1,aoki2,aoki3}, the massless pion corresponds to
the inverse of the divergent correlation length at the second order phase
transition point. Since  the effective theory which describes this phase 
transition is some 4 dimensional scalar(pion) theory with rather complicated 
interactions\footnote{Indeed our WChPT is an approximation
of this effective theory.},
the phase transition has the mean-field critical exponent with possible
log-corrections. In particular the pion mass, the inverse of the correlation 
length, should behaves near the critical point as
\beqa
m_\pi^2 &=& C m_R \left\{\log\left(\frac{m_R}{D}\right) \right\}^{\nu^\prime} + \cdots ,
\eeqa
where $\cdots$ represent less singular contributions.
If we expand
\beqa
\left\{\log\left(\frac{m_R}{D}\right) \right\}^{\nu^\prime}
&=&\left\{\log\left(\frac{\Lambda_0^2}{A D}\right)
+ \log\left(\frac{A m_R}{\Lambda_0^2}\right)\right\}^{\nu^\prime} 
= X^{\nu^\prime}\sum_{n=0}^\infty \frac{\nu^\prime !}{ (\nu^\prime -n)! n!}
\left(\frac{Y}{X}\right)^n
\nn \\
&=& X^{\nu^\prime} \left( 1 +\nu^\prime \frac{Y}{X} + \cdots \right)
\eeqa
where
\beqa
X &=& \log\left(\frac{\Lambda_0^2}{A D}\right) \\
Y &=& \log\left(\frac{A m_R}{\Lambda_0^2}\right),
\eeqa
the formula at the next-to-leading order in WChPT, eq.(\ref{eq:fit-mpi}),
is recovered, with the identification that
\beqa
\frac{\nu^\prime}{X} &=& \frac{a^2 w_0}{32\pi^2f^2}, \qquad
C X^{\nu^\prime} = A .
\eeqa 
To determine $\nu^\prime$ and $X$ separately, the explicit calculation in the WChPT 
at 2-loop or more orders is necessary. 
This will be considered in future investigations.

We have finally obtained the following resummed formulae for $m_\pi^2$ and
$m_{\rm AWI}$: 
\beqa
m_\pi^2 &=& A m_R \left\{\log \left(\frac{m_R}{\Lambda_0}\right)
\right\}^{\frac{a^2 w_0}{32\pi^2 f^2}} 
\left[ 1+ \frac{m_R A+m_R a w_1}{32\pi^2 f^2}
\log \left(\frac{A m_R}{\Lambda^2}\right)
 \right]  \\
\label{eq:fit-mpi.rs}
m_{\rm AWI}&=& A_0 m_R \left\{\log \left(\frac{m_R}{\Lambda_0}\right)
\right\}^{\frac{a^2 w_0}{32\pi^2 f^2}} 
\left[ 1+ \frac{m_R a w_1^{\rm AWI}
}{32\pi^2 f^2}\log  \left(\frac{A m_R}{\Lambda_{\rm AWI}^2}\right)
\right] ,
\label{eq:fit-mawi.rs}
\eeqa
where $A$, $\Lambda_0$ and $\omega_0$ may be different from those
in eqs.(\ref{eq:fit-mpi},\ref{eq:fit-mawi}).
It is better to use these formulae instead of the previous ones, 
eqs.(\ref{eq:fit-mpi},\ref{eq:fit-mawi}), in future investigations.
Eq.(\ref{eq:ratio}) remains the same.

As a trial, we use these formulae with $A_0=1$, $\Lambda_{\rm AWI}=\Lambda$
and $\Lambda_0$=1 GeV, in order to fit $m_\pi^2$ and $m_{\rm AWI}$ simultaneously,
at each $a$. The quality of the fit is as good as the previous one, and
the fitting parameters are compiled in the end of table~\ref{tab:WChPT}.
In addition,
the next-to-leading contribution, the second term in eq.(\ref{eq:fit-mpi.rs}),
vanishes toward $m_R=0$ 
as shown in Fig.~\ref{fig:next}, where 
$R$(WChPT) in the previous subsection, which is now modified as
\beqa
R(\mbox{WChPT, resum}) &=&
\frac{m_R A+m_R a w_1}{32\pi^2 f^2}
\log \left(\frac{A m_R}{\Lambda^2}\right),
\eeqa
are plotted at $\beta=$ 1.8, 1.95, 2.1 and 2.2 .

\section{Conclusions and Discussions}
\label{sec:conclusion}
In this paper we have derived the effective chiral Lagrangian which includes
the $a^2$ effect of the Wilson-type quark action in the case of the $N_f=2$
degenerate quarks. Using this effective Lagrangian
the quark mass($m_R$) dependences of $m_\pi^2$, $m_{\rm AWI}$ and 
$f_\pi$ have been calculated at the 1-loop level.
We then have simultaneously fitted $m_\pi^2$ and $m_{\rm AWI}$,
obtained by the CP-PACS collaboration for $N_f=2$ full QCD simulations,
using the WChPT formula, and have found that the data are 
consistently described. We have attempted the continuum extrapolation
of the WChPT formula.

Comparing to the standard ChPT, 
several distinct features such as the additive mass renormalization,  
$O(a)$ corrections to the chiral log($m_R\log m_R$) term,
a more singular term($\log m_R$) generated by $O(a^2)$ contributions
and the presence of both $m_R\log m_R$ and $\log m_R$ terms in  $m_{\rm AWI}$,
leads to the success for the WChPT formula to describe the CP-PACS data.
Although an ambiguity for the definition of $K_c$ caused by the additive mass
renormalization can be avoided by the use of $m_{\rm AWI}$, the last feature,
the existence of both $m_R\log m_R$ and $\log m_R$ terms in $m_{\rm AWI}$,
makes the WChPT formula different from the ChPT's.
The large $O(a)$ correction to $m_R\log m_R$ term
plays an essential role to describe the actual data,
though more or less others have some contributions.
We have also derived the formula after resumming $\log m_R$ terms,
using the fact  that the mean-field critical exponent
receives the log-correction.

Because of the limitation of available data, our WChPT analysis is far from
complete. Therefore it is important to refine the analysis by taking
the correlation between $m_\pi^2$ and $m_{\rm AWI}$ into account and
including $f_\pi$ data in the simultaneous fit, in order to
establish the validity of the WChPT.
Reanalyses of other full QCD data  have to be done of course.
It is also urgent to derive the WChPT formula for other cases\cite{BRS} such
as the quench/partially quench cases, the $N_f=3$ non-degenerate case,
vector mesons and baryons, heavy-light mesons.

Once the validity of the WChPT to describe lattice QCD data
is established, instead of thinking that the quark masses in the current full 
QCD simulations are too heavy for the ChPT to apply,
we may say that some (but not all) of lattice data are well described by
the (Wilson) chiral perturbation theory, by which
errors associated with the chiral extrapolation may be well controlled
\cite{ks}.

\section*{Acknowledgments}
I would like to thank Drs. O. B\"ar, N.~Ishizuka and A.~Ukawa
for useful discussions.
This work is supported in part by the Grants-in-Aid for
Scientific Research from the Ministry of Education, 
Culture, Sports, Science and Technology.
(Nos. 13135204, 14046202, 15204015, 15540251).

\bibliography{basename of .bib file}

\begin{table}[h]
\caption{Parameters of $N_f=2$ full QCD simulations by the CP-PACS collaboration
\protect{\cite{cppacs}}. The scale $a$ is fixed by $m_\rho =$ 768.4 MeV.}
\label{tab:cppacs}
\begin{center}
\begin{tabular}{lllllll}
\hline
$\beta$ & $L^3\times T$ & $c_{\rm SW}$ & $a$ [fm] & $a^{-1}$ [GeV] & $L a$ [fm] &
$m_\pi/m_\rho$ \\
\hline
1.80 & $12^3\times 24$ & 1.60 & 0.2150(22) & 0.9178(94) & 2.580(26) &
0.55$\sim$ 0.81 \\
1.95 & $16^3\times 32$ & 1.53 & 0.1555(17) & 1.269(14) & 2.489(27) & 
0.58$\sim$ 0.80 \\
2.10 & $24^3\times 48$ & 1.47 & 0.1076(13) & 1.834(22) & 2.583(31) & 
0.58$\sim$ 0.81 \\
2.20 & $24^3\times 48$ & 1.44 & 0.0865(33) & 2.281(87) & 2.076(79) & 
0.63$\sim$ 0.80 \\
\hline
\end{tabular}
\end{center}
\end{table}

\begin{table}[h]
\caption{Parameters of the WChPT fit at each $\beta$.}
\label{tab:WChPT}
\begin{center}
\begin{tabular}{l|clllll|c}
\hline
$\beta$ &$K_c$ &$A$[GeV] & $\Lambda$ [GeV] &$w_1a$[GeV] 
&$w_0a^2$[GeV$^2$] & $w_1^{\rm AWI}a$ [GeV]& $\chi^2$/dof \\
\hline
1.80 & 0.147761(15)& 5.114(28)& 0.079(19)& -5.525(64)& 0.206(22)&-0.560(74)& 0.3\\
1.95 & 0.142160(19)& 5.377(33)& 0.193(51)& -5.162(74)& 0.241(42)&-0.457(118)&0.3\\
2.10 & 0.139110(12)& 5.807(14)& 0.694(20)& -5.24(18) & 0.417(50)&-1.15(27) & 0.2\\
2.20 & 0.137691(23)& 5.669(71)& 0.128(88)& -5.15(20) & 0.039(16)&-0.22(39)& 0.7\\
\hline
& \multicolumn{6}{c|}{resummed WChPT} &\\
\hline
1.8 & 0.147562(15)& 5.111(29)& 0.067(12) & -4.862(46) & 0.787(21)
    & 0.124(15) & 1.5 \\
1.95& 0.142009(7) & 5.366(23) & 0.132(15) & -4.538(52) & 0.624(18)
    & 0.310(32) & 0.3 \\
2.1 & 0.138959(13) & 5.535(47) & 0.131(71) & -4.79(14)& 0.280(37)
    & 0.181(49) & 1.2 \\
2.2 & 0.137657(36) & 5.789(106) & 0.391(82) & -4.63(12)& 0.201(95)
    & 0.195(96) & 0.8 \\
\hline
\end{tabular}
\end{center}
\end{table}

\begin{table}[h] 
\caption{Continuum extrapolation of the WChPT fit parameters.
(a) $m_\pi^2$ and $m_{\rm AWI}$ are fitted as a function of $m_R$
at each $a$. Then parameters are fitted as a function of $a$.
(b) $m_\pi^2/m_{\rm AWI}$ are fitted as a function of $m_R$ and $a$
}
\label{tab:global} 
\begin{center} 
\begin{tabular}{c|cccc|c|ccc}
\hline 
\multicolumn{5}{l|}{(a)} & \multicolumn{4}{|l}{ (b)\qquad $\chi^2$/dof=1.3}\\
\hline
$X$  &$d_0(X)$ &$d_1(X)$ &$d_2(X)$ &$\chi^2$/dof &
$X$  &$d_0(X)$ &$d_1(X)$ &$d_2(X)$  \\
\hline
$K_c$&-0.2127(10) &-0.008300(55) & 0.000787(31) & 3.6 &
$A$  & 8.087(97) & -1.002(29) & 0.2672(29) \\
$w_0$& 0.202(17)  & 0 & 0 & 20 &
$\Lambda$& 1.196(35) &-0.8404(58) & 0 \\ 
$w_1^{\rm AWI}$ &-0.549(61) & 0 & 0 & 3.4 &
$\Delta w_1$ & -1.62(25)& 0 & 0  \\
& & & & &
$A_0$ & -0.590(47) &0 & 0\\
\hline
\end{tabular}
\end{center}
\end{table}

\clearpage

\end{document}